\documentclass[a4paper]{article}
\usepackage{epsfig,amsmath,amsfonts,amssymb,setspace,multirow,textcomp}
\usepackage[T1]{fontenc}
\makeatletter
\renewcommand{\@evenfoot}{\hfil \thepage \hfil}
\renewcommand{\@oddfoot}{\hfil \thepage \hfil}
\makeatother

\renewenvironment{thebibliography}[1]{\begin{oldthebibliography}{#1}\setlength{\parskip}{0ex}\setlength{\itemsep}{0ex}}{\end{oldthebibliography}}

\begin{document}
\fontsize{11}{11}\selectfont 
\title{Unidentified aerial phenomena. Observations of variable objects }
\author{Boris Zhilyaev$^{1}$, David Tcheng$^{2}$, Vladimir Petukhov$^{1}$}
%
\date{\vspace*{-6ex}}
\maketitle
\begin{center} {\small $^{1}$$Main \,Astronomical \, Observatory, NAS \, of \, Ukraine, Zabalotnoho \,27, 03680, Kyiv, Ukraine$}\\
{\small $^{2}$The University of Illinois, 1200 West Harrison St. Chicago, Illinois 60607}\\

{\small bzhi40@gmail.com}

\end{center}

\begin{abstract}
NASA commissioned a research team to study Unidentified Aerial Phenomena (UAP). The Main Astronomical Observatory of NAS of Ukraine conducts an independent study of UAP. A research team from San Diego also decided to conduct a study of UAP. Observations of events that cannot scientifically be identified as known natural phenomena established the existence of the UAP. However, their nature remains unclear. For UAP observations, we used a meteor station installed in Kyiv. We also present a new methodology for detecting UAPs using the latest smartphone technology in San Diego. We inform about discovering a new type of UAP.


\end{abstract}

\section{Introduction}

The Main Astronomical Observatory of NAS of Ukraine conducts an independent study of unidentified phenomena in the atmosphere. A research team from San Diego also decided to conduct a study of UAP. 

Unidentified anomalous, air, and space objects are deeply concealed phenomena. The main feature of the UAP is its high speed.
Ordinary photo and video recordings will not capture the UAP. To detect UAP, we need to fine-tune (tuning) the equipment: shutter speed, frame rate, and dynamic range.

According to our data, there are two types of UAP, which we conventionally call: (1) Cosmics, and (2) Phantoms. We note that Cosmics are luminous objects, brighter than the background of the sky. Phantoms are dark objects, with a contrast, according to our data, to several per cent. Both types of UAPs exhibit high movement speeds. Their detection is a difficult experimental problem. 

The results of previous UAP study are published in \cite{Zhilyaev2022}. Here we present our recent results.

Kyiv astronomers have identified three groups of objects (1) a group of bright spinning objects, (2) a group of bright structured objects and (3) a group of dark flying objects. Monitoring of the daytime sky led to the detection of bright and dark objects, moving at a speed from about 1M to 16M and sizes from about 20 to 100 meters. The detection of these objects, according to Ukrainian astronomers, is an experimental fact.

The results were confirmed most recently from ample observations in San Diego revealing the shape and behaviour patterns we describe here. 

We report the discovery of a new type of UAP, which we tentatively call "blinkers". They demonstrate regular bursts of brightness with a frequency in the range up to 20 Hz and a duration of hundredths of a second. The critical feature of blinkers is the drop in brightness to near zero. 


\section{OBSERVATIONS AND DATA PROCESSING}

For UAP observations, we used a meteor station installed in Kyiv. The station has an ASI 294 Pro camera and lens with a focal length of 28 and 50 mm.  ASI 294 Pro camera has a FOV of up to 9.7 deg, a pixel size of 34.1 and 19.1 arc second, and a frame rate of up to 120 fps.

The SharpCap 4.0 program was used for data recording. Observations of objects were carried out in the daytime sky. Frames were recorded in the .ser format with 14 bits. 

For San Diego observations, two modern smartphones were used, the Samsung s23 Ultra and the Pixel 6 Pro.  
The Samsung was used for high-speed single-camera video captures with a maximum rate of 960 frames per second.  The spatial resolution of the smartphone is  1080 x 1920 in precise 8-bit color per pixel and a FOV of up to 85 x 48 deg.

Video observations were made in Escondido California, with a clear sky view.  Observations were made during daylight.  

Our best observations were made with the Samsung s23 Ultra using the "Super Slow Motion Mode". 
We recorded with two smartphone cameras (Samsung and Google) simultaneously at 240 fps. This allows us to filter out low flying known species of bird and insects.

For both single and multiple camera setups, we made many recordings.  Each Samsung high speed (960 fps) capture resulted in a 17 s video at 30 fps.  The dual camera recordings at 240 fps were manually started and stopped with a target length of 1 minute and more. Any video with any recognizable object movement was saved for further review and categorization. 

We have written Matlab code for adding false color to RGB videos to bring out the contrast.  

\begin{figure}[!h]
\centering
\begin{minipage}[t]{.45\linewidth}
\centering
\epsfig{file = 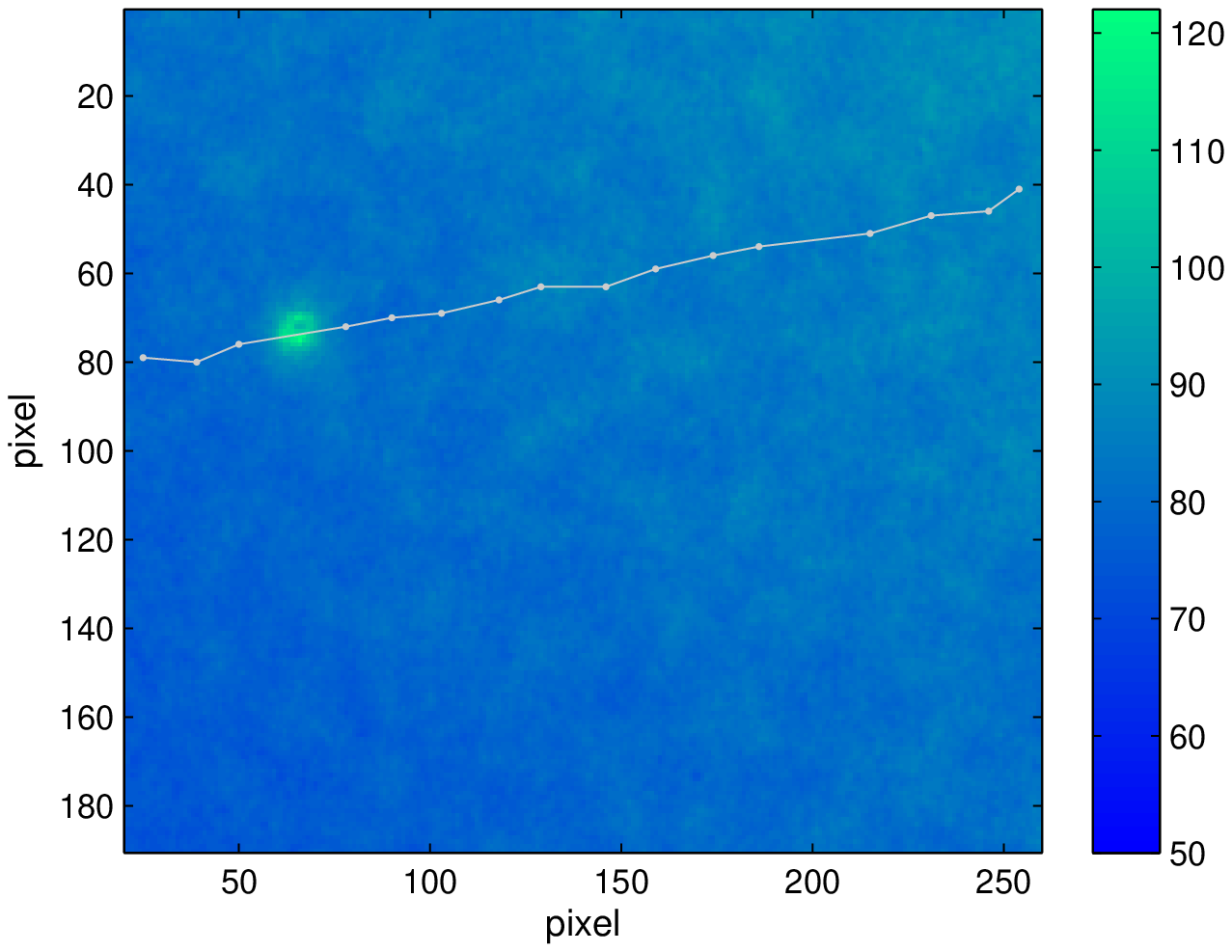,width = 1.05\linewidth} \caption{UAP track (Kyiv on June 30, 2022).}
\end{minipage}
\hfill
\begin{minipage}[t]{.45\linewidth} 
\centering
\epsfig{file = 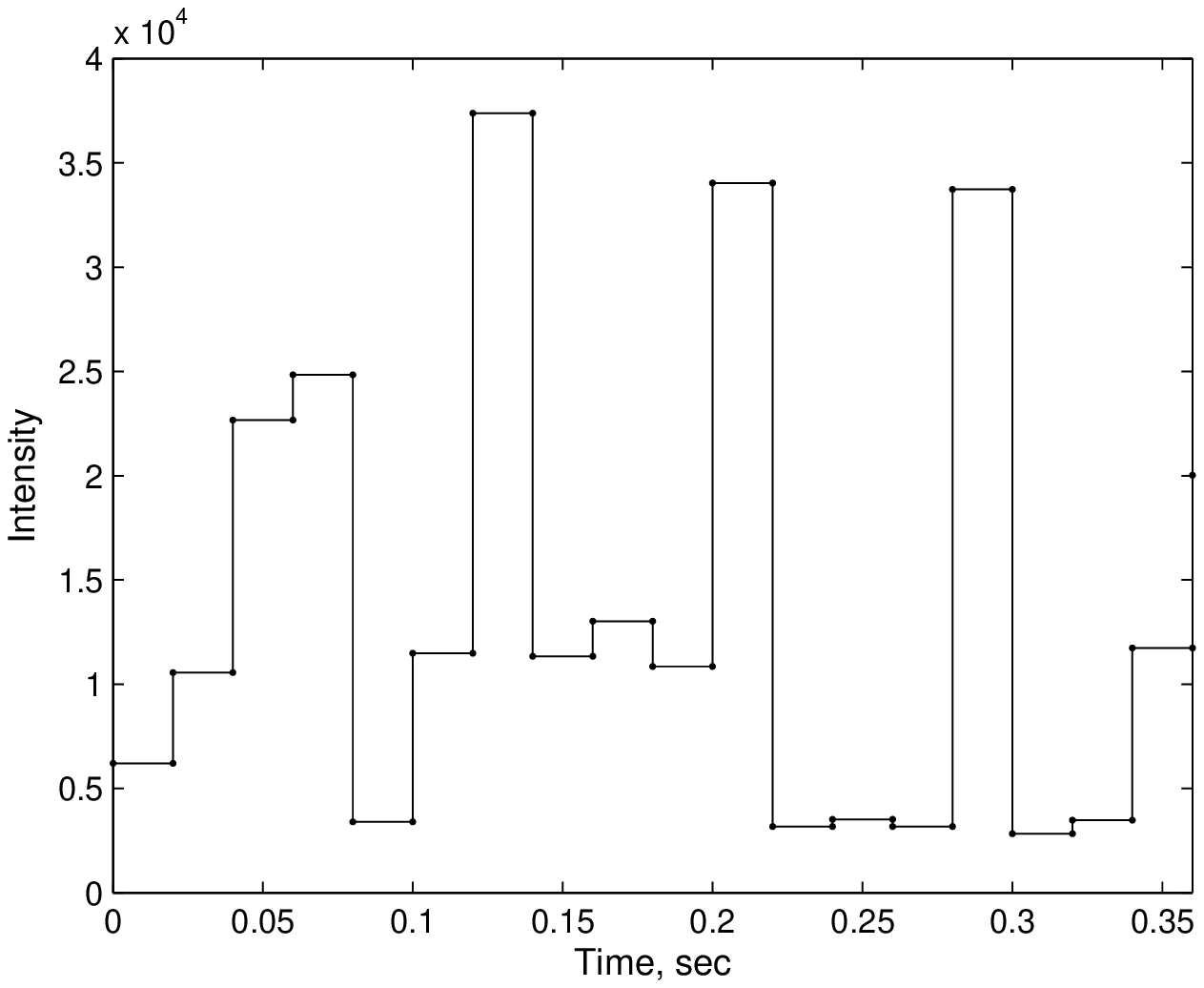,width = 1.05\linewidth} \caption{The light curve of UAP on June 30, 2022. The object flashes for no more than two-hundredth of a second at an average of 10 times per second.}\label{fig2}
\end{minipage}
\end{figure}
\section{Results}
Observations in Kyiv on June 30, 2022, and February 12, 2023 revealed UAP objects of a particular type. This group of bright objects of large angular sizes has clearly defined structural features. Two of them have almost identical characteristics. 

A feature of this group is regular flashes of objects. In addition, the intensity of the glow drops to near zero.

Observation start times are (1) UTC: 06/30/2022 07:22:08; (2) UTC: 12/02/2023 11:02:58. Objects were observed in the daytime sky at the zenith.

Object number 1 on  June 30, 2022 in Fig. 1 crosses a frame of 2.2 degrees for 0.40 sec with 50 frames per second with 1 ms exposure. It demonstrates a speed of 5.5 degrees per second and periodicity with a frequency of about 10 Hz (Fig. 2). Some frames in Fig. 3 show the intensity of the glow drops to 7 times or about 2 stellar magnitudes.

Fig. 3 shows its size of about 25 pixels (14.2 arc minutes, about half the size of the Moon), which indicates the final dimensions. Its contrast is about 22\%. If we assume that it is at a distance of 1 km, its size will be about 7 meters, if at a distance of 4 km, then 28 meters. In the latter case, its speed will be about 380 m/s (about 1M).

Figs 4, 5 and 6 show object number 2 on February 12, 2023 with 59.4 frames per second.

These two objects were observed in the central part of Kyiv with an interval of 8 months, in summer and winter, respectively. The latter is important since insects, the main source of interference in observations of bright objects, are absent during winter. Figures 3 and 6 clearly defined structural features, they have almost identical characteristics.

Figs. 2 and 5 show the objects flash for no more than two-hundredth of a second at an average of 10 and 20
times per second. The special feature of the objects is the intensity of the glow drops to almost zero. It
is natural to assume that a bright object shines by reflected sunlight. But this is not compatible with
the intensity drop to zero.

Both objects demonstrate a toroidal image structure with bright islands. The second object also exhibits rotation. Image details rotated approximately 90 degrees in 0.13 seconds.

Figs. 7, 8 and 9 show the UAP observed over San Diego on April 04, 2023.  The object demonstrates strong variations of intensity (Fig. 8). The object flashes for no more than three-hundredths of a second. The next feature of the object is the intensity of the glow drops to zero. Notably, the object's light curve shows a repeating signal.
\begin{figure}[!h]
\centering
\begin{minipage}[t]{.45\linewidth}
\centering
\epsfig{file = 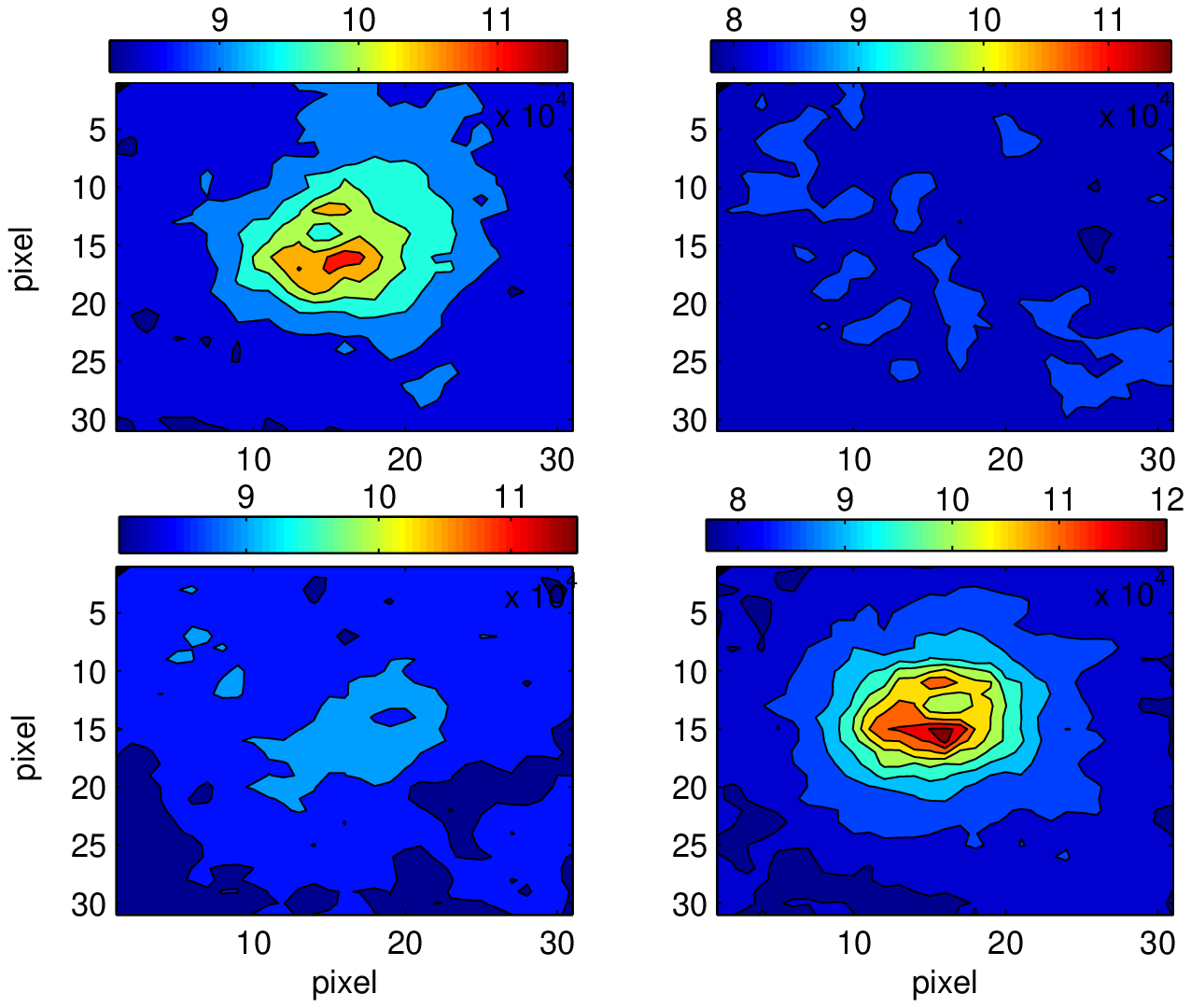,width = 1.05\linewidth} \caption{The sequence of UAP images on June 30, 2022.}
\end{minipage}
\hfill
\begin{minipage}[t]{.45\linewidth} 
\centering
\epsfig{file = 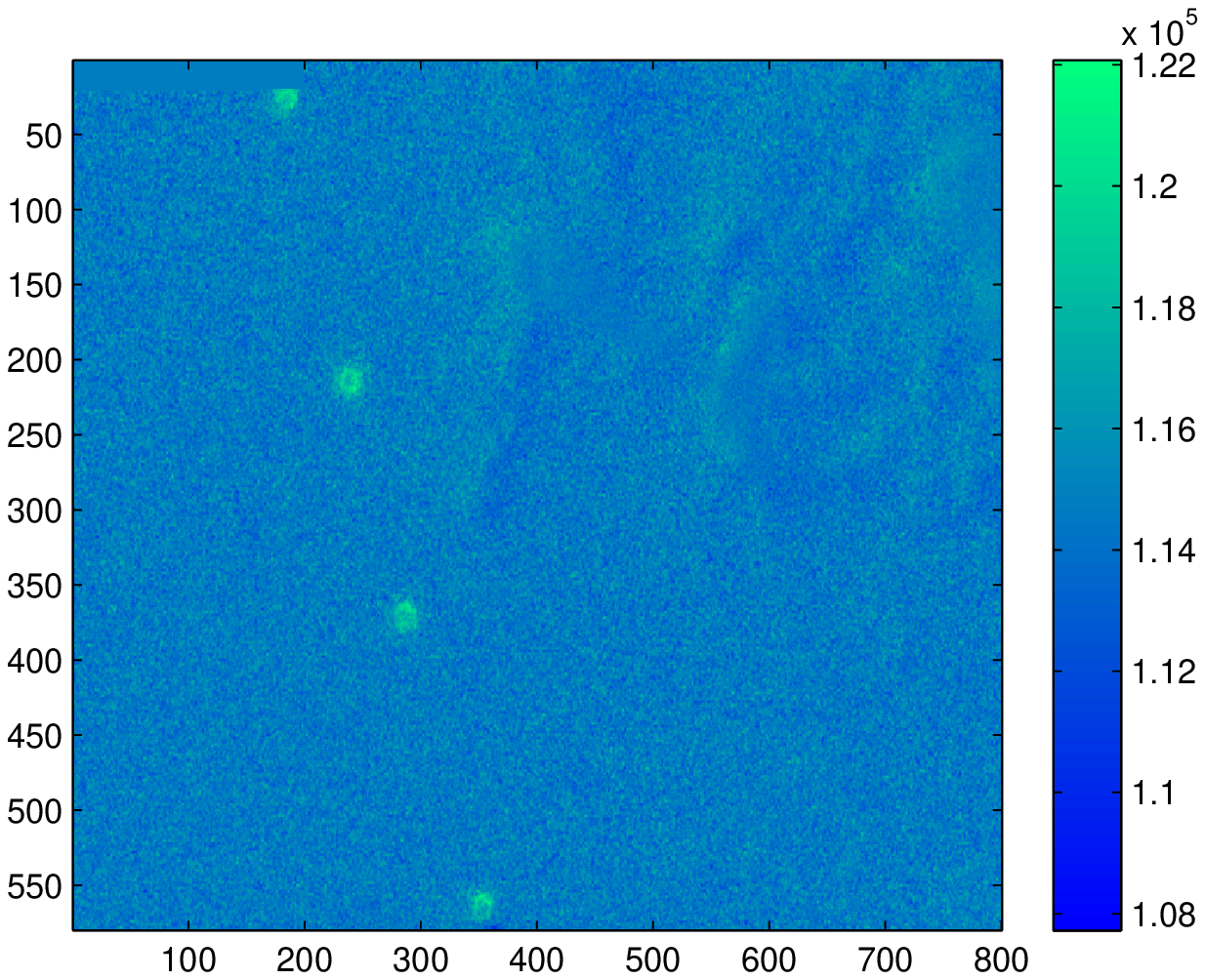 ,width = 1.05\linewidth} \caption{UAP track (Kyiv 2023).}\label{fig2}
\end{minipage}
\end{figure}
\begin{figure}[!h]
\centering
\begin{minipage}[t]{.45\linewidth}
\centering
\epsfig{file = 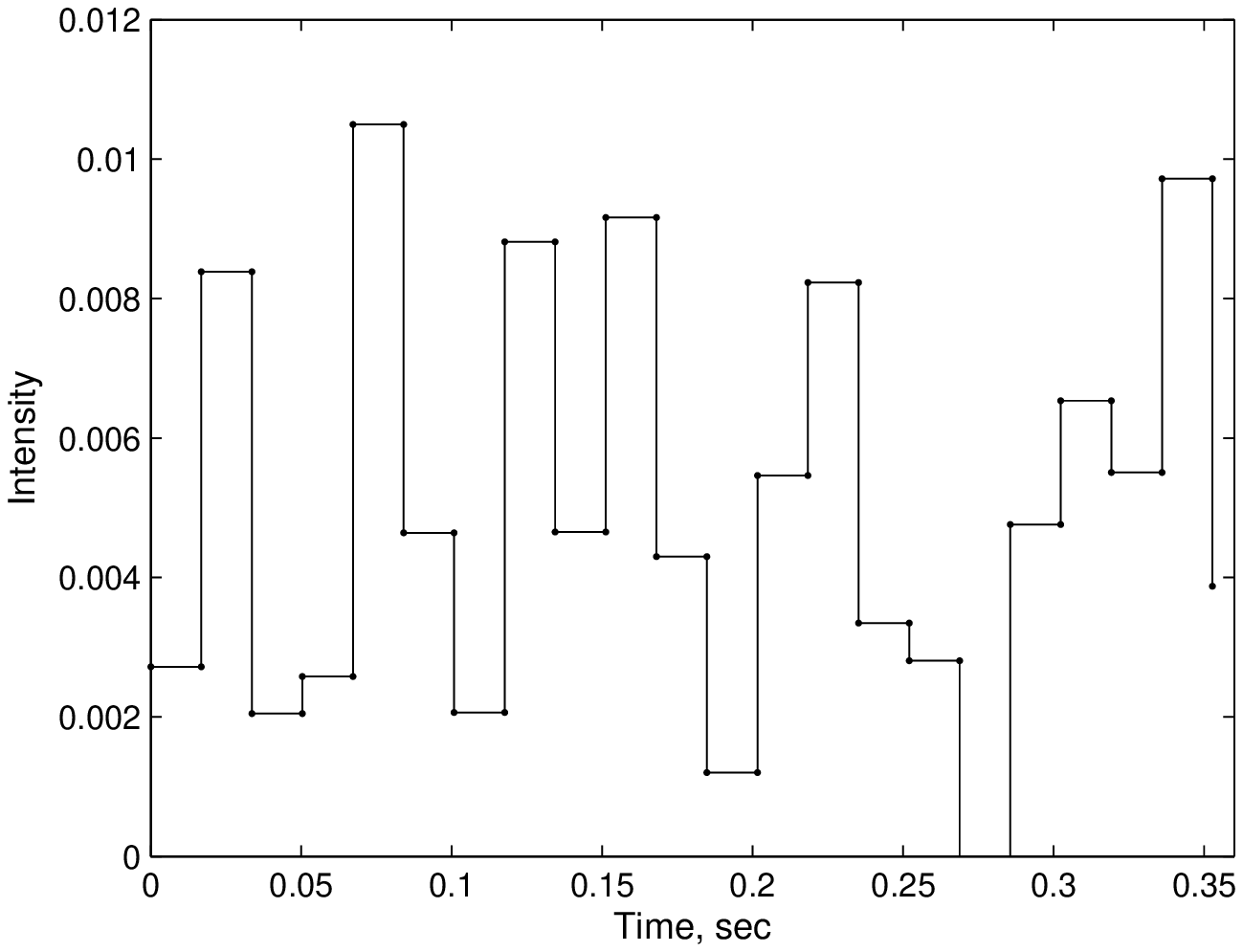,width = 1.05\linewidth} \caption{The light curve of the UAP on March 14 2023, Kyiv.}
\end{minipage}
\hfill
\begin{minipage}[t]{.45\linewidth} 
\centering
\epsfig{file = 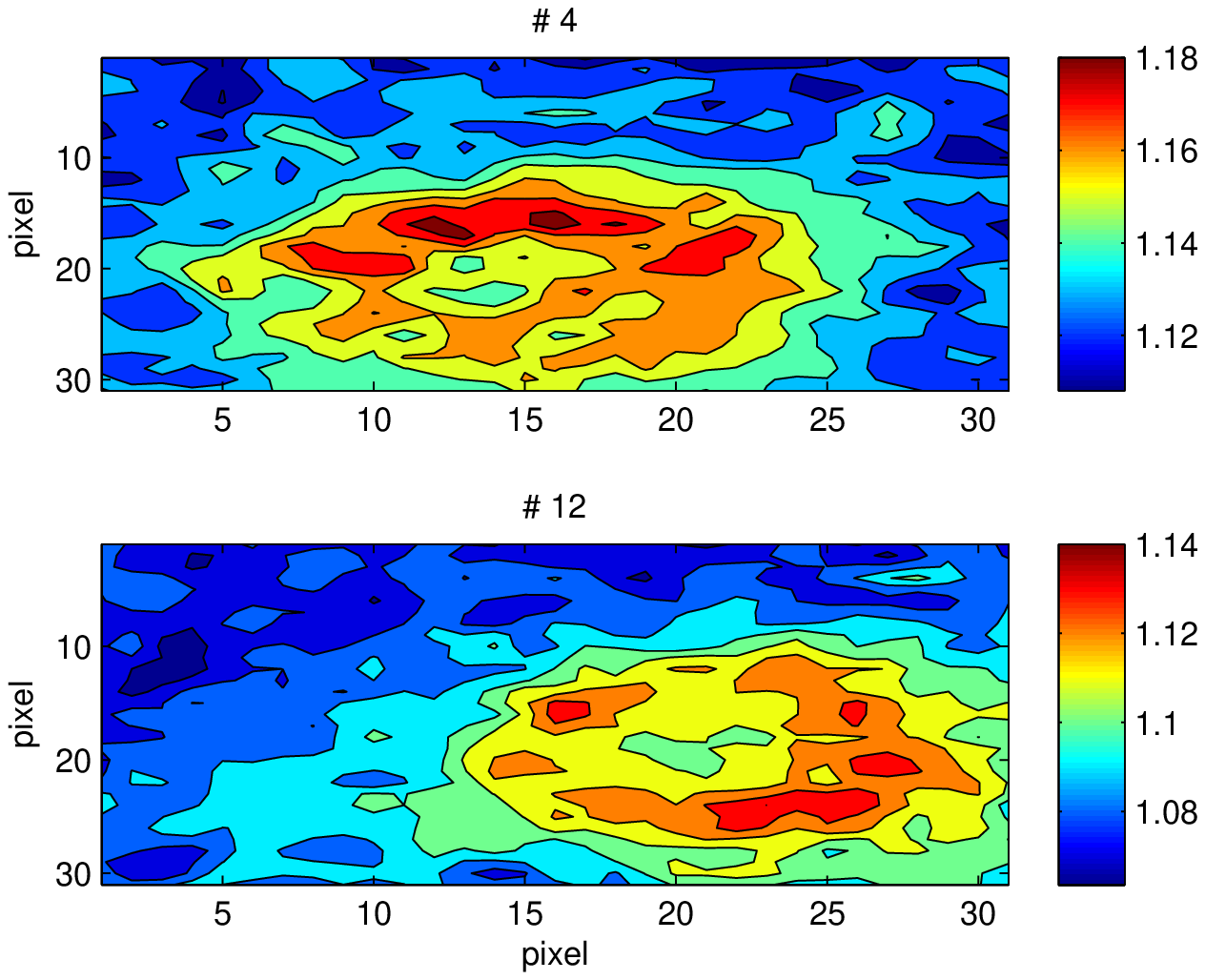,width = 1.05\linewidth} \caption{The images of UAP on March 14 2023, Kyiv.}\label{fig2}
\end{minipage}
\end{figure}

\begin{figure}[!h]
\centering
\begin{minipage}[t]{.45\linewidth}
\centering
\epsfig{file = 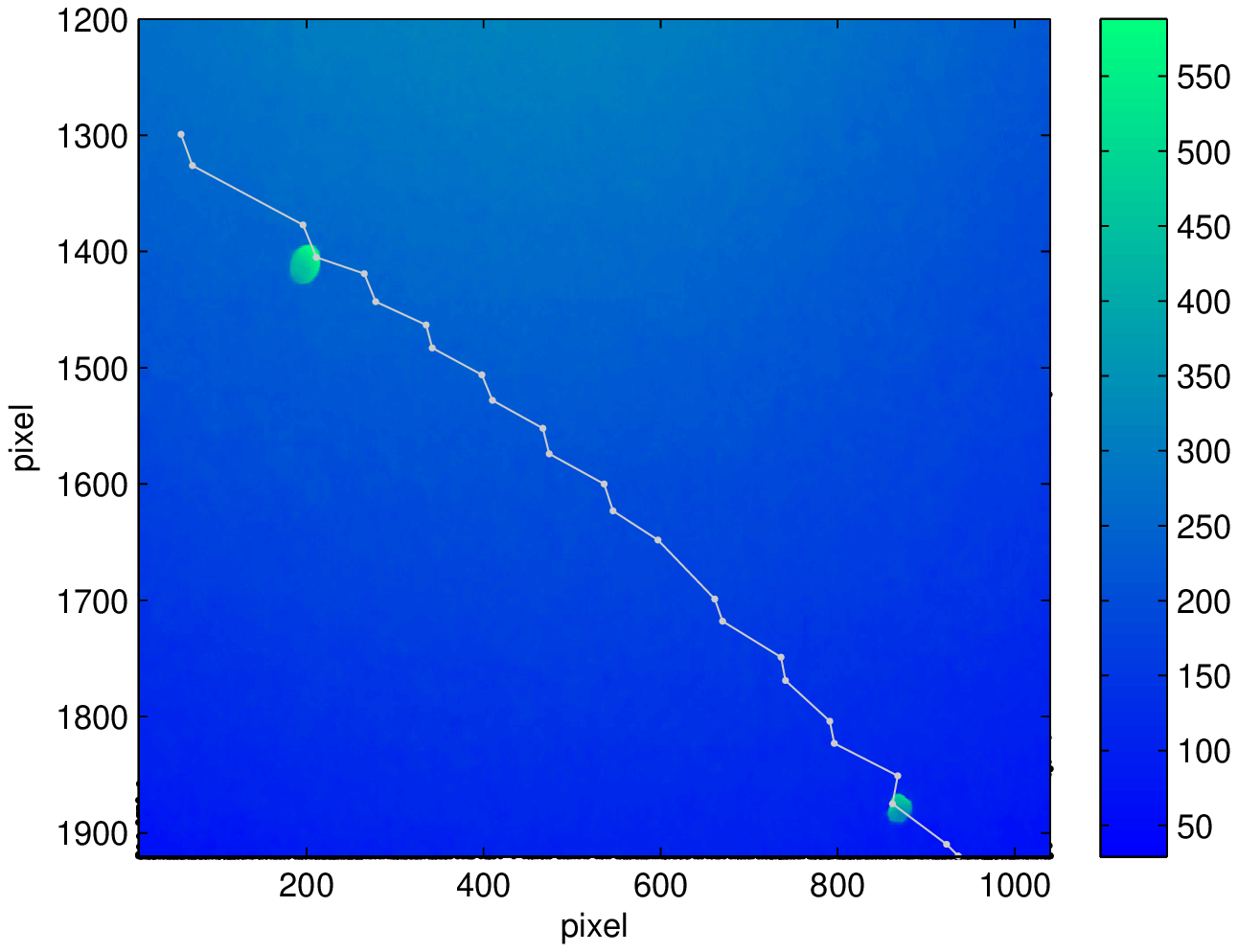,width = 1.05\linewidth} \caption{UAP track on April 4, 2023,\,(San Diego).}
\end{minipage}
\hfill
\begin{minipage}[t]{.45\linewidth} 
\centering
\epsfig{file = 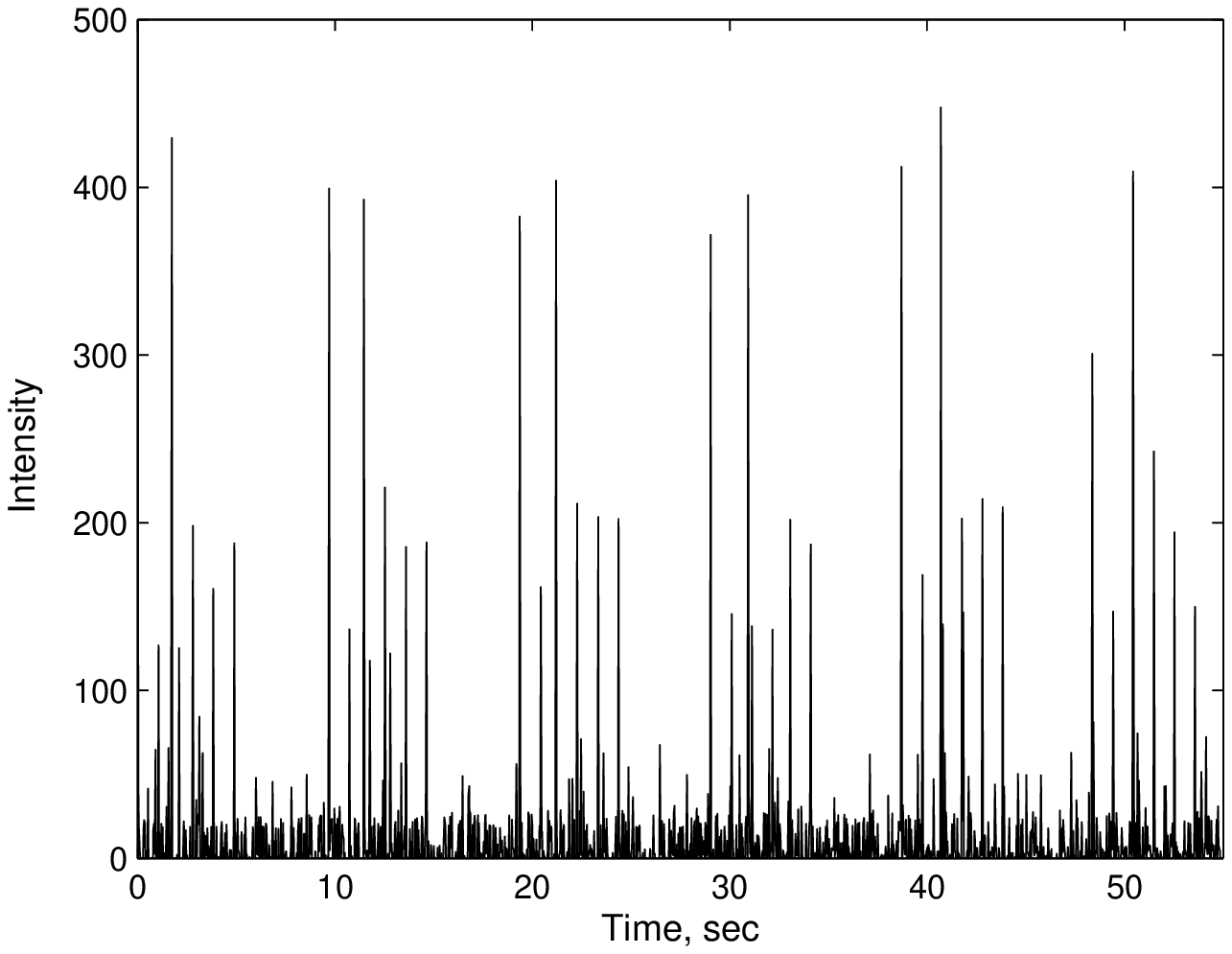 ,width = 1.05\linewidth} \caption{UAP light curve on April 4, 2023,\,(San Diego).}\label{fig2}
\end{minipage}
\end{figure}

\begin{figure}
\centering
\resizebox{0.5\hsize}{!}{\includegraphics[angle=000]{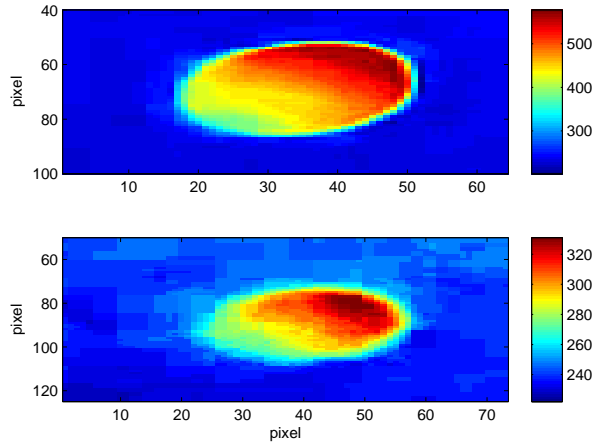}}%
\caption{UAP images on April 4, 2023,\,(San Diego) for large amplitude double pulse and triple pulse.} \label{figure: track1_1.eps}
\end{figure}

The object in Fig. 7 crosses a frame of 38 degrees for 135 sec with 30 frames per second. It demonstrates a speed of 0.3 degrees per second and periodicity with a frequency of about 0.104 Hz (Fig. 8). The object's size is about 60 arc minutes (two sizes of the Moon).

There are three features in the object's light curve. (1) Large amplitude double pulses. (2) Two triple pulses with half the amplitude. (3) All three features are cyclically repeated.

The double pulse repeats with a period of 9.62 $\pm $ 0.20 sec over an observation interval of about 135 sec. The interval between the peaks of the double pulse changes monotonically from 2 to 2.5 sec.

The interval between peaks in triple pulses remains constant and equal to 1.03 sec. The triple flashes are synchronized with the start of the twin flash.

Noteworthy is the fact that the pulse repetition periods have values close to 1 and 10 seconds.

The San Diego object (Fig. 9) has almost characteristics like an ellipse. The shape and size of the ellipsoidal figure remain unchanged. Fig. 9 shows images of objects for large amplitude double pulse and triple pulse with half the amplitude. Its angular size is four times greater than the size of the Kyiv objects. This may be due to the distance to the object.

\section{Discussion}
A professional ASI camera in Kyiv and two modern smartphones used, the Samsung s23 Ultra and the Pixel 6 Pro, in San Diego observations, discovered a new type of UAP. The advantage of ASI cameras is high metrology and modern smartphones in higher FOV and fps.

\newpage

We inform about discovering a new type of UAP, which we conventionally call "blinkers". A feature of this object is regular flashes with a duration of hundredths of a second. In addition, the intensity of their glow drops to almost zero.

The waveform of blinkers can be characterized by a duty cycle, i.e. the percentage of the ratio of pulse duration, or pulse width to the total period of the waveform. Its duty cycle ranges from about 1 to 10 per cent. This means that blinkers can be invisible ninety-nine per cent of the time.

%


\end{document}